\documentclass[9pt,twocolumn,letterpaper]{extarticle}

\usepackage[utf8]{inputenc}
\usepackage[T1]{fontenc}
\usepackage{textcomp}
\usepackage[letterpaper,margin=0.76in,top=0.74in,bottom=0.74in,columnsep=0.25in]{geometry}
\usepackage{newtxtext}
\usepackage[cmintegrals]{newtxmath}
\usepackage[numbers,sort&compress]{natbib}
\usepackage{graphicx}
\usepackage{amsmath}
\usepackage{url}
\usepackage{microtype}
\usepackage[percent]{overpic}
\usepackage{caption}
\graphicspath{{figures/}}
\DeclareGraphicsExtensions{.pdf,.png,.jpg,.jpeg}

\begin{document}

\twocolumn[{
\begin{center}
{\Large\bfseries Chlorophyll-a Fluorescence-Based Optical Temperature Sensor Using Food-Grade Chlorella Extract\par}
\vspace{9pt}
{\small Sinei Felix Gonçalves Junior\textsuperscript{1},
Giovanna G. S. de Miranda\textsuperscript{1},
Leomar S. Marques\textsuperscript{1},
Flavio A. M. Marques\textsuperscript{1},
Alexandre A. C. Cotta\textsuperscript{1},
Jefferson E. Tsuchida\textsuperscript{1},
Julio C. Ugucioni\textsuperscript{1},
Lidia O. Rosa\textsuperscript{2},
Eric Fujiwara\textsuperscript{3},
Cristiano M. B. Cordeiro\textsuperscript{2},
and Jonas H. Osório\textsuperscript{1}\par}
\vspace{6pt}
{\footnotesize
\textsuperscript{1} Laboratório Multiusuário de Óptica e Fotônica, LaMOF, Departamento de Física, Universidade Federal de Lavras, Lavras, Brazil\par
\textsuperscript{2} Instituto de Física Gleb Wataghin, Universidade Estadual de Campinas, UNICAMP, Campinas, Brazil\par
\textsuperscript{3} Faculdade de Engenharia Mecânica, Universidade Estadual de Campinas, UNICAMP, Campinas, Brazil\par}
\end{center}
\vspace{8pt}
\noindent

\vspace{0pt}
\small
\noindent\textbf{Abstract.} This letter reports the development of an optical temperature sensor based on the fluorescence of an extract obtained from \textit{Chlorella}, a unicellular microalga commercially available as food-grade biomass. Using a straightforward ethanolic extraction route, the sensing platform exploits the characteristic emission of chlorophyll-\textit{a} as the optical transducing parameter. The operating principle relies on thermal fluorescence quenching, which promotes a decrease in emission intensity with rising temperature. In the temperature range from $28.6$ to $43.0~^{\circ}\text{C}$, the sensor exhibited a linear response ($R^2 = 0.9963$) with a sensitivity of $(-8.9 \pm 0.2) \times 10^{-3}~^{\circ}\text{C}^{-1}$. 
Dynamic testing over consecutive heating--cooling cycles confirmed the sensor operation under sub-degree thermal oscillations and enabled the characterization of the system's photobleaching drift ($3.65\%\cdot\text{h}^{-1}$). The results reported herein demonstrate the feasibility of exploiting sustainable, bio-derived microalgal pigments as functional photoactive materials for optical thermometry.
\par\vspace{14pt}
}]

\section{Introduction}

Optical temperature sensors based on luminescence thermometry attract considerable attention due to their non-invasive nature, immunity to electromagnetic interference, and suitability for remote monitoring in harsh or delicate environments \cite{Britesfluorescence}. In particular, fluorophore-based thermal sensors operating via thermal fluorescence quenching offer fast response times and high spatial resolution \cite{brites2012thermometry}. However, conventional luminescent thermometers typically incorporate synthetic dyes (such as rhodamines) \cite{liu2014multifluorescent}, semiconductor quantum dots \cite{birchall2022inkjet}, or lanthanide-doped phosphors \cite{brites2019lanthanide}, which frequently involve hazardous precursors, cytotoxicity, or complex and high-cost synthesis routes. Consequently, expanding the library of sustainable, non-toxic, and bio-derived photoactive materials represents an important step forward for green photonics and environmentally benign sensing platforms.

Photosynthetic pigments, prominently chlorophylls, present remarkable light-absorption and emission properties in the visible and near-infrared regions \cite{pareek2017chlorophylls}. While chlorophyll fluorescence has been widely used \textit{in vivo} as a non-destructive diagnostic tool to assess plant physiology \cite{baker2008chlorophyll} and fruit maturation \cite{cerovic2008new}, isolated chlorophylls have gained increasing traction as functional materials in artificial optoelectronic and photonic devices. Recent demonstrations include chlorophyll-sensitized solar cells \cite{wuryanti2021solar}, organic phototransistors \cite{yu2019photoelectricity}, light-driven soft actuators \cite{wang2025curling}, and chemical sensors for selective heavy-metal ion detection \cite{wu2018copper} and water turbidity analysis \cite{attivissimo2015portable}.

Despite these advances, the temperature dependence of chlorophyll fluorescence has predominantly been treated from a physiological or photobiological standpoint rather than being explored as an optical transducing mechanism for thermal sensing, so that natural photosynthetic pigments extracted from renewable biomass remain largely unexplored for this purpose. In this context, microalgae such as \textit{Chlorella} represent a particularly advantageous source as they are commercially available, and provide chlorophyll yields that can be readily extracted using benign or food-grade protocols \cite{cuaresma2011photobioreactors, ferreira2023stabilization}.

In this framework, here, we report the development of an optical temperature sensor based on the thermal fluorescence quenching of chlorophyll-\textit{a} extracted from commercially available, food-grade \textit{Chlorella} biomass. Using a single-step ethanolic extraction, the system operates across the temperature range of $28.6$ to $43.0~^{\circ}\text{C}$ with a linear response ($R^2 = 0.9963$) and a sensitivity of $ (-8.9 \pm 0.2) \times 10^{-3} \text{ }^{\circ}\text{C}^{-1}$. Furthermore, dynamic thermal cycling allowed evaluating the sensor operation under sub-degree thermal oscillations and characterizing the system photobleaching drift. The results reported herein demonstrate the feasibility of employing unrefined microalgal extracts as sustainable, accessible photoactive elements for optical thermometry.

\section{METHODOLOGY}

\subsection{Sample Preparation}

In this work, pigments were extracted from commercially sourced food-grade \textit{Chlorella} biomass (powder form), known for its high content of chlorophyll \cite{cuaresma2011photobioreactors, ferreira2023stabilization}. Extraction followed a straightforward solvent-assisted protocol, as described in the following: 5.0~g of microalgal biomass was dispersed into 20.0~mL of absolute ethanol (99.8\% purity, Fig.~1(a)) and subjected to continuous magnetic stirring for 20~min at room temperature (Fig.~1(b)) to promote cellular membrane permeabilization and pigment dissolution \cite{ventura2017extraction}. The resulting suspension was filtered through filter paper combined with a cotton plug to remove suspended microparticles and cell debris, thereby reducing light-scattering losses during optical interrogation (Fig.~1(c)). An aliquot of 750~$\mu$L of the concentrated extract was subsequently diluted in 50.0~g of deionized water to obtain the optical working solution (Fig.~1(d)). The final solution exhibited the characteristic green coloration under ambient illumination (Fig.~1(e)) and red emission under near-UV optical excitation (Fig.~1(f)), corresponding to the typical fluorescence of chlorophyll-\textit{a}.

The absorbance spectrum of the diluted extract is shown in Fig.~1(g). The spectral profile exhibits the characteristic features of chlorophyll-\textit{a}, displaying the Soret band in the blue region and the Q-band in the red region, along with minimal absorption across the green spectral window. Beyond confirming the preservation of the pigment following the extraction protocol, the strong absorption within the Soret band provides the rationale for selecting a laser diode emitting at $405$~nm as the optical source for fluorescence excitation.

\subsection{Optical Setup and Sensing Principle}
The experimental arrangement for optical characterization and temperature sensing is illustrated in Fig.~2(a). The sample solution was contained in a glass cuvette positioned in direct contact with a Peltier thermoelectric module. A calibrated thermocouple (0.1 ºC resolution) immersed directly into the liquid sample recorded the reference temperature throughout the tests. Here, the thermocouple was held in place by a custom 3D-printed cap that replicated the original cuvette lid, featuring a central through-hole to conveniently position the probe into the liquid sample.

Optical excitation was provided by a continuous-wave diode laser emitting at $\lambda = 405$~nm with an optical output power of 4.5~mW (Thorlabs CPS405). The pump beam was focused into the sample using an objective lens ($4\times$, L1). The emitted fluorescence was collected at a transmission geometry by a second objective lens ($5\times$, L2), passed through a long-pass optical filter (Thorlabs FELH0500, cut-on wavelength at 500~nm) to suppress residual pump scattering, and focused via a third objective ($10\times$, L3) into the input fiber of a CCD spectrometer (Thorlabs CCS200/M) operating with an integration time of 2.0~s. The spectrometer used in this work features a spectral measurement range of 200–1000 nm and a nominal spectral resolution of < 2 nm FWHM at 633 nm.

The transducing mechanism of the sensor reported herein relies on the thermal quenching of chlorophyll-\textit{a} fluorescence \cite{zscheilc1943fluorescence}. As temperature rises, the quantum yield of the radiative emission is reduced, hence lowering the measured fluorescence signal around the dominant peak at $\lambda \approx 680$~nm. The performance of the sensor is evaluated in terms of its corresponding sensitivity, taking into account the variation of the fluorescence intensity as a function of temperature changes.

\begin{figure}[ht!]
    \centering
    \begin{overpic}[percent,width=0.99\columnwidth]{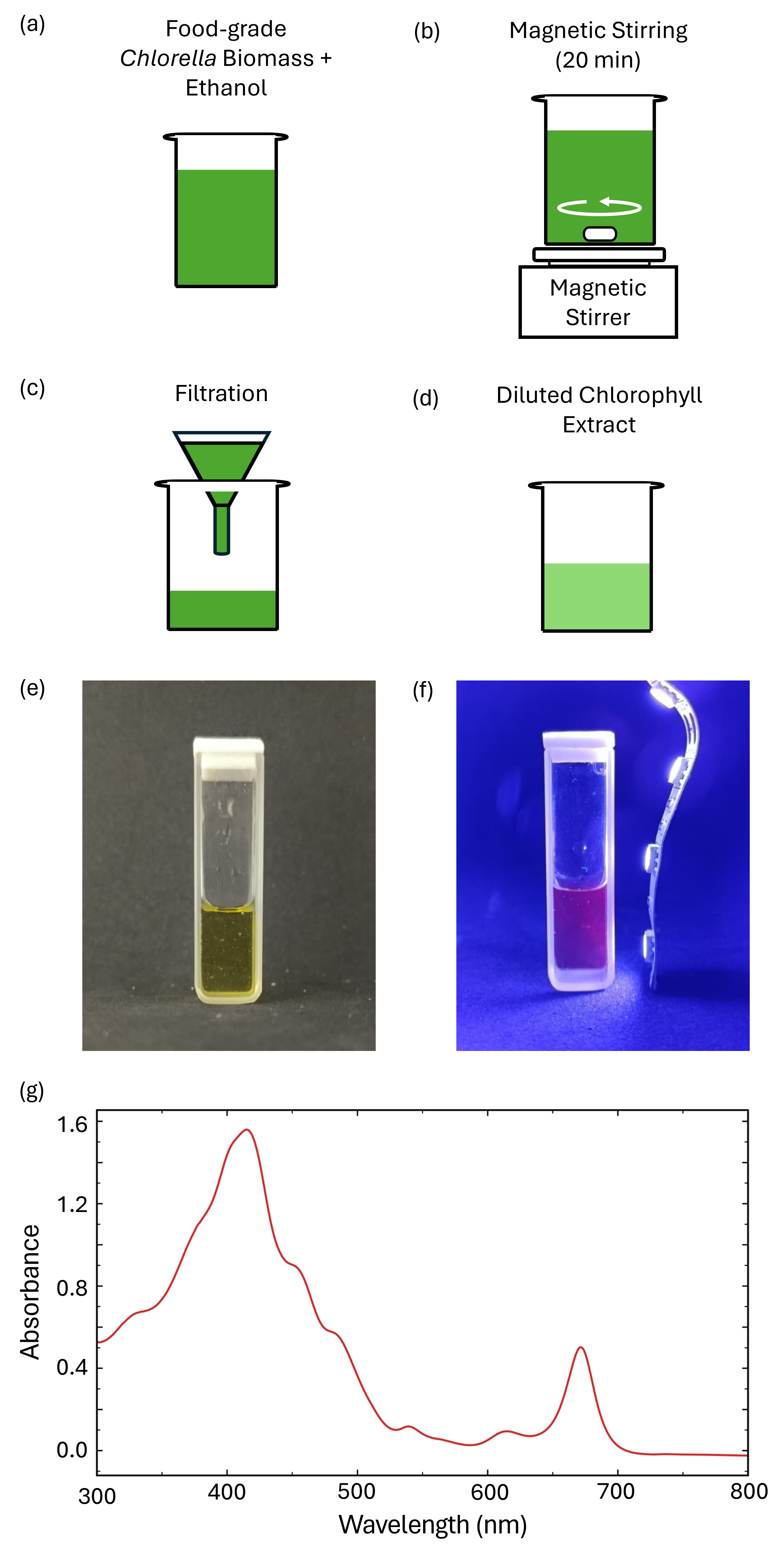}
    \end{overpic}
    \caption{Preparation, visual inspection and absorbance spectrum of the chlorophyll extract from food-grade \textit{Chlorella} biomass. (a) \textit{Chlorella} dispersion in ethanol. (b) Magnetic stirring (20 min). (c) Gravity filtration. (d) Diluted extract in deionized water. (e) Cuvette inspection under ambient lighting. (f) Cuvette under UV excitation displaying characteristic red fluorescence. (g) Absorbance spectrum of the diluted extract in the 300–800 nm range.}
    \label{fig:chlorophyll_extraction}
\end{figure}

\begin{figure}[ht!]
    \centering

    \begin{overpic}[percent,width=0.8\columnwidth]{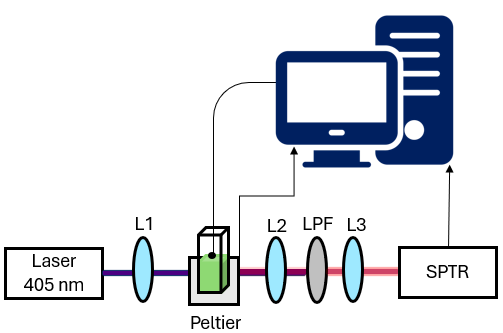}
    
    \end{overpic}

    \vspace{1mm}

    \caption{Schematic representation of the experimental setup comprising excitation laser, lenses (L1--L3), temperature-controlled cuvette mount, long-pass optical filter (LPF), and spectrometer (SPTR).}
    \label{fig:optical-setup}
\end{figure}

\section{RESULTS AND DISCUSSION}

\subsection{Sensor Calibration and Thermal Response}
The optical response of the food-grade \textit{Chlorella} extract to temperature changes was investigated by tracking the fluorescence emission spectra from $28.6$ to $43.0~^{\circ}\text{C}$. This upper limit was chosen to encompass relevant physiological and environmental monitoring ranges, while ensuring that the signal variation is governed by reversible thermal quenching, below the temperature threshold where thermal degradation of chlorophyll takes place in liquid media (> 50–60 °C) \cite{zscheilc1943fluorescence}. 

As shown in Fig.~3(a), the emission spectra display a dominant peak centered at $\lambda \approx 680$~nm, along with a secondary shoulder extending up to $740$~nm \cite{zscheilc1943fluorescence}. Increasing the temperature induces a steady reduction across the entire emission profile due to thermal fluorescence quenching, with no evident spectral shifts. The peak fluorescence intensity decreased monotonically over the investigated range, representing an overall drop of $20.5\%$.

To quantify sensor performance, the peak fluorescence intensity was extracted for each thermal step and plotted as a function of temperature, as depicted in Fig.~3(b) (the error bars represent standard deviation over 6 repetitions). The calibration curve displays a linear behavior yielding a coefficient of determination of $R^2 = 0.9963$. From the linear regression slope, a sensitivity of $ (-8.9 \pm 0.2) \times 10^{-3} \text{ }^{\circ}\text{C}^{-1}$ was determined. The linearity and evaluated thermal response confirm the viability of the unrefined chlorophyll extract as an optical transducing element for temperature monitoring.

\begin{figure}[ht!]
    \centering
    \begin{overpic}[percent,width=0.9\columnwidth]{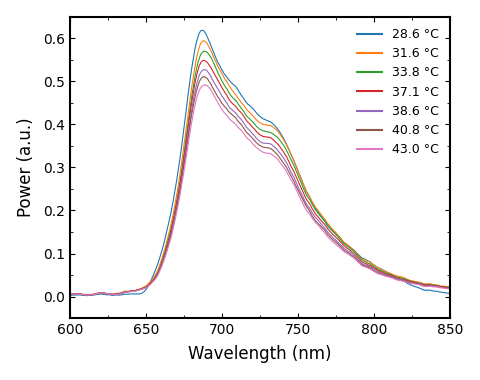}
        \put(0.5,75){\footnotesize\text{(a)}}
    \end{overpic}
    \begin{overpic}[percent,width=0.9\columnwidth]{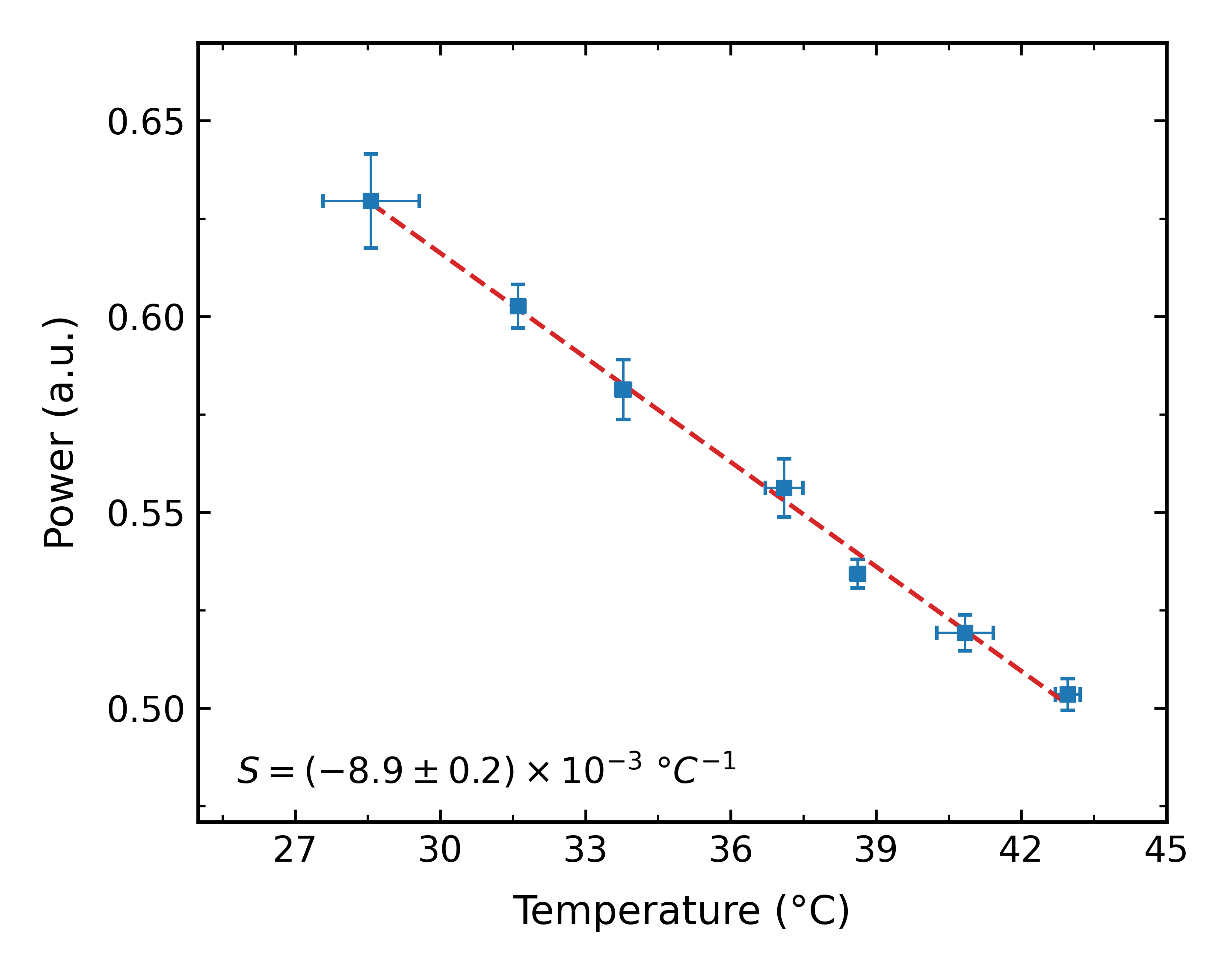}
        \put(0.5,75){\footnotesize\text{(b)}}
    \end{overpic}

    \caption{(a) Chlorophyll-\textit{a} fluorescence spectra as a function of temperature from 28.6 to 43.0~$^{\circ}$C. (b) Peak fluorescence intensity as a function of temperature along with the linear calibration curve.}
\label{fig:calibration}
\end{figure}

\subsection{Thermal Cycling and Photobleaching Analysis}

To evaluate sensor response under continuous operation, real-time measurements were performed over a 60-min period under quasi-periodic room temperature variations. Fig.~4(a) displays the reference temperature recorded by the thermocouple, while the optical readout extracted at the peak emission wavelength is depicted in Fig.~4(b). As expected, the measured fluorescence signal increases during cooling and decreases during heating.

A direct comparison reveals a systematic baseline decay in the raw optical response (Fig.~4(b)) due to the induction of a photobleaching drift of the chlorophyll fluorophores. The baseline decay (dashed line in Fig.~4(b)) followed a linear trend ($R^2 = 0.9891$), corresponding to a photodegradation rate of $3.65\%\cdot\text{h}^{-1}$, which was fitted through the consecutive local minima of the thermal cycles. Here, we understand that this moderate measured photobleaching rate under continuous laser exposure  is primarily governed by mass-transport dynamics within the liquid cuvette. Since optical interrogation is confined to the focused waist of the pump laser beam, molecular diffusion and thermally induced convection continuously replenish unbleached fluorophores from the bulk liquid reservoir into the excitation volume. Furthermore, because total photodegradation remained low over the 60-min window, the baseline drift followed a linear behavior, enabling straightforward baseline compensation, as we will describe in the following.

Thus, to isolate the thermal transduction component, the linear photobleaching baseline was subtracted from the optical signal, yielding the drift-compensated response shown in Fig.~4(c). By applying the calibration sensitivity, the optical signal was converted into equivalent temperature variations ($\Delta T_c$), as presented in Fig.~4(d). This procedure hence allowed for recovering the temperature variations from the optical measurements. 

To quantify the tracking accuracy of the reconstructed thermal signal, the absolute discrepancy between $\Delta T_{c}$ and the thermocouple reference $\Delta T$ was analyzed at the local extrema across all four cycles. The average absolute deviation at these turning points was $0.05~^{\circ}\mathrm{C}$, with individual differences displaying a maximum of $0.08~^{\circ}\mathrm{C}$. This low residual error demonstrates that, following the photobleaching baseline compensation, the optical sensor achieves a conservative dynamic temperature resolution better than $0.1~^{\circ}\mathrm{C}$, being capable of faithfully resolving sub-degree thermal oscillations 

\begin{figure}[ht!]
    \centering

    \begin{overpic}[percent,width=0.95\columnwidth]{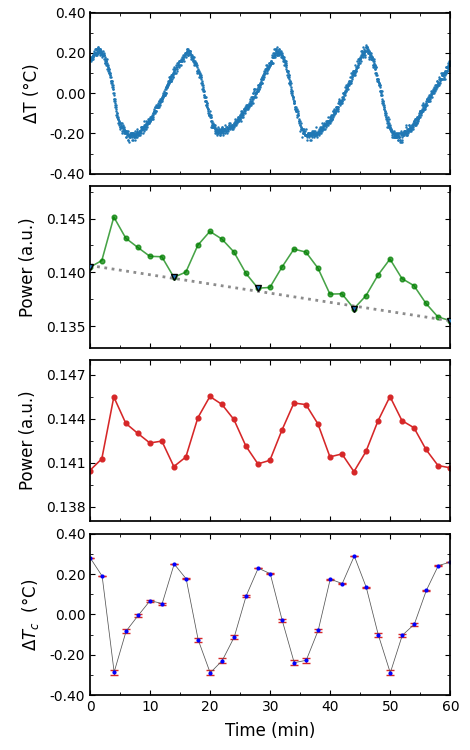}
        \put(-2.5,98){\footnotesize\text{(a)}},
        \put(-2.5,73){\footnotesize\text{(b)}}
        \put(-2.5,50){\footnotesize\text{(c)}}
        \put(-2.5,26){\footnotesize\text{(d)}}
    \end{overpic}
    \caption{Dynamic thermal cycling analysis over 60 min. (a) Reference temperature recorded by the thermocouple. (b) Raw optical readout showing thermal oscillations superimposed on the linear photobleaching drift (dashed line). (c) Drift-compensated fluorescence intensity. (d) Optically reconstructed temperature variations ($\Delta T_c$).}

    \label{fig:temporal-analysis}
\end{figure}

\section{CONCLUSION}
In this letter, we demonstrated an optical temperature sensing approach based on the thermal fluorescence quenching of chlorophyll-\textit{a} extracted from food-grade \textit{Chlorella} biomass. Prepared via a simple ethanolic extraction and aqueous dilution route, the sensing platform exhibited a linear response ($R^2 = 0.9963$) across the $28.6$ to $43.0~^{\circ}\text{C}$ range, with a sensitivity of $ (-8.9 \pm 0.2) \times 10^{-3} \text{ }^{\circ}\text{C}^{-1}$. Dynamic thermal cycling revealed that baseline drift associated with photobleaching ($3.65\%\cdot\text{h}^{-1}$) can be compensated, enabling the optical readout to track sub-degree thermal fluctuations. These findings highlight the potential of pigments from microalgal biomass as functional, accessible, and environmentally benign photoactive materials, paving the way for future developments in bio-derived luminescence thermometry, including optofluidic and fiber-optic probes.

\section{ACKNOWLEDGEMENTS}
The authors thank FAPEMIG (RED-00046-23, APQ-00197-24, APQ-01618-25, APQ-07070-26) and CNPq (407961/2022-4, 305024/2023-0, 402723/2024-4, 409174/2024-6, 301312/2026-5), and FAPESP (2024/02995-4, 2024/00998-6) for financial support.

\bibliographystyle{unsrtnat}
\bibliography{references}

@incollection{pareek2017chlorophylls,
  author    = {Pareek, Sunil and Sagar, Narashans Alok and Sharma, Sunil and Kumar, Vinay and Agarwal, Tripti and Gonz{\'a}lez-Aguilar, Gustavo A. and Yahia, Elhadi M.},
  title     = {Chlorophylls: Chemistry and Biological Functions},
  booktitle = {Fruit and Vegetable Phytochemicals: Chemistry and Human Health},
  edition   = {2},
  editor    = {Yahia, Elhadi M.},
  publisher = {Wiley},
  pages     = {269--284},
  year      = {2017},
  doi       = {10.1002/9781119158042.ch14},
  isbn      = {9781119157946}
}

@article{yu2019photoelectricity,
  author  = {Yu, Yu and Zhang, Yating and Jin, Lufan and Chen, Zhiliang and Li, Yifan and Li, Qingyan and Cao, Mingxuan and Che, Yong and Dai, Haitao and Yang, Jing and Yao, J. Q.},
  title   = {Photoelectricity and Thermoelectricity in Organic Chlorophyll Phototransistors},
  journal = {Organic Electronics},
  volume  = {65},
  pages   = {381--385},
  month   = feb,
  year    = {2019},
  doi     = {10.1016/j.orgel.2018.11.039}
}

@article{wang2025curling,
  author  = {Wang, Hao and Jiang, Jiwei and Sun, Yunyu and Zhang, Hui and Song, Bo and Dong, Bin},
  title   = {Fast Light-Driven Curling Actuation Based on Agar-Chlorophyll Film},
  journal = {ChemistrySelect},
  volume  = {10},
  number  = {17},
  pages   = {e202501266},
  year    = {2025},
  doi     = {10.1002/slct.202501266}
}

@article{attivissimo2015portable,
  author  = {Attivissimo, Filippo and Guarnieri Cal{\`o} Carducci, Carlo and Lanzolla, Anna Maria Lucia and Massaro, Alessandro and Vadrucci, Maria Rosaria},
  title   = {A Portable Optical Sensor for Sea Quality Monitoring},
  journal = {IEEE Sensors Journal},
  volume  = {15},
  number  = {1},
  pages   = {146--153},
  month   = jan,
  year    = {2015},
  doi     = {10.1109/JSEN.2014.2340437}
}

@article{wu2018copper,
  author  = {Wu, Meng-Jer and Hu, Hsiu-Hao and Siao, Chen-Zih and Liao, Yu-Ming and Chen, Jyun-Han and Li, Ming-Yang and Lin, Tai-Yuan and Chen, Yang-Fang},
  title   = {All Organic Label-Like Copper({II}) Ions Fluorescent Film Sensors with High Sensitivity and Stretchability},
  journal = {ACS Sensors},
  volume  = {3},
  number  = {1},
  pages   = {99--105},
  year    = {2018},
  doi     = {10.1021/acssensors.7b00635}
}

@article{wuryanti2021solar,
  author  = {Wuryanti, Sri},
  title   = {The Performance of Solar Cells Using Chlorophyll Dye from {Syzygium paniculatum}},
  journal = {Clean Energy},
  volume  = {5},
  number  = {3},
  pages   = {433--440},
  year    = {2021},
  doi     = {10.1093/ce/zkab022}
}

@article{liu2014multifluorescent,
  author  = {Liu, Hengjun and Maruyama, Hisataka and Masuda, Taisuke and Honda, Ayae and Arai, Fumihito},
  title   = {Multi-Fluorescent Micro-Sensor for Accurate Measurement of {pH} and Temperature Variations in Micro-Environments},
  journal = {Sensors and Actuators B: Chemical},
  volume  = {203},
  pages   = {54--62},
  year    = {2014},
  doi     = {10.1016/j.snb.2014.06.079}
}

@article{birchall2022inkjet,
  author  = {Birchall, Liesbeth and Foerster, Aleksandra and Rance, Graham A. and Terry, Anna and Wildman, Ricky D. and Tuck, Christopher J.},
  title   = {An Inkjet-Printable Fluorescent Thermal Sensor Based on {CdSe/ZnS} Quantum Dots Immobilised in a Silicone Matrix},
  journal = {Sensors and Actuators A: Physical},
  volume  = {347},
  pages   = {113977},
  year    = {2022},
  doi     = {10.1016/j.sna.2022.113977}
}

@article{cuaresma2011photobioreactors,
  author  = {Cuaresma, Mar{\'i}a and Janssen, Marcel and V{\'i}lchez, Carlos and Wijffels, Ren{\'e} H.},
  title   = {Horizontal or Vertical Photobioreactors? How to Improve Microalgae Photosynthetic Efficiency},
  journal = {Bioresource Technology},
  volume  = {102},
  number  = {8},
  pages   = {5129--5137},
  year    = {2011},
  doi     = {10.1016/j.biortech.2011.01.078}
}

@article{ferreira2023stabilization,
  author  = {Ferreira, Andreia S. and Pereira, Liliana and Canfora, Feliciana and Silva, Tiago H. and Coimbra, Manuel A. and Nunes, Cl{\'a}udia},
  title   = {Stabilization of Natural Pigments in Ethanolic Solutions for Food Applications: The Case Study of {Chlorella vulgaris}},
  journal = {Molecules},
  volume  = {28},
  number  = {1},
  pages   = {408},
  year    = {2023},
  doi     = {10.3390/molecules28010408}
}

@incollection{ventura2017extraction,
  author    = {Ventura, S. P. M. and Nobre, B. P. and Ertekin, F. and Hayes, M. and Garcia-Vaquero, M. and Vieira, F. and Koc, M. and Gouveia, L. and Aires-Barros, M. R. and Palavra, A. M. F.},
  title     = {Extraction of Value-Added Compounds from Microalgae},
  booktitle = {Microalgae-Based Biofuels and Bioproducts: From Feedstock Cultivation to End-Products},
  editor    = {Gonzalez-Fernandez, Cristina and Mu{\~n}oz, Ra{\'u}l},
  chapter   = {19},
  pages     = {461--483},
  publisher = {Woodhead Publishing},
  year      = {2017},
  doi       = {10.1016/B978-0-08-101023-5.00019-4},
  isbn      = {9780081010235}
}

@article{zscheilc1943fluorescence,
  author  = {Zscheilc, F. P. and Harris, D. G.},
  title   = {Studies on the Fluorescence of Chlorophyll. The Effects of Concentration, Temperature, and Solvent},
  journal = {The Journal of Physical Chemistry},
  volume  = {47},
  number  = {9},
  pages   = {623--637},
  year    = {1943},
  doi     = {10.1021/j150432a001}
}

@article{Britesfluorescence,
  author  = {Brites, C. D. S. and Marin, R. and Suta, M. and Carneiro Neto, A. N. and Ximendes, E. and Jaque, D. and Carlos, L. D.},
  title   = {Spotlight on Luminescence Thermometry: Basics, Challenges, and Cutting-Edge Applications},
  journal = {Advanced Materials},
  volume  = {35},
  number  = {36},
  pages   = {2302749},
  year    = {2023},
  doi     = {10.1002/adma.202302749}
}

@article{brites2012thermometry,
  author    = {Brites, Carlos D. S. and Lima, Patricia P. and Silva, Nuno J. O. and Mill{\'a}n, {\'A}ngel and Amaral, V{\'i}tor S. and Palacio, Fernando and Carlos, Lu{\'i}s D.},
  title     = {Thermometry at the nanoscale},
  journal   = {Nanoscale},
  volume    = {4},
  number    = {16},
  pages     = {4799--4829},
  year      = {2012},
  publisher = {Royal Society of Chemistry},
  doi       = {10.1039/c2nr30663h}
}

@article{brites2019lanthanide,
  author    = {Brites, Carlos D. S. and Balabhadra, Sangeetha and Carlos, Lu{\'\i}s D.},
  title     = {Lanthanide-based thermometers: At the cutting-edge of luminescence thermometry},
  journal   = {Advanced Optical Materials},
  volume    = {7},
  number    = {5},
  pages     = {1801239},
  year      = {2019},
  publisher = {Wiley Online Library},
  doi       = {10.1002/adom.201801239}
}

@article{baker2008chlorophyll,
  author  = {Baker, Neil R.},
  title   = {Chlorophyll fluorescence: a probe of photosynthesis in vivo},
  journal = {Annual Review of Plant Biology},
  volume  = {59},
  pages   = {89--113},
  year    = {2008},
  doi     = {10.1146/annurev.arplant.59.032607.092759}
}

@article{cerovic2008new,
  author  = {Cerovic, Zoran G. and Moise, Nathalie and Agati, Giovanni and Latouche, Gwendal and Ben Ghozlen, Nabil and Meyer, Sylvie},
  title   = {New portable optical sensors for the assessment of winegrape phenolic maturity based on berry fluorescence},
  journal = {Journal of Food Composition and Analysis},
  volume  = {21},
  number  = {8},
  pages   = {650--654},
  year    = {2008},
  doi     = {10.1016/j.jfca.2008.03.012}
}

\end{document}